\documentstyle[aps, multicol]{revtex}

\def\etal{\emph{et al.}\ }

\input BoxedEPS
\SetepsfEPSFSpecial
\HideDisplacementBoxes

\begin{document}

\title{Magnetoresistance of proximity coupled Au wires}
\author{D.\ A.\ Dikin, M.\ J.\
 Black and V.\ Chandrasekhar}
\address{ Department of Physics and Astronomy, Northwestern University, Evanston, IL 60208, USA \\ }

\maketitle

\begin{abstract} We report measurements of the magnetoresistance (MR) of narrow Au wires coupled to a superconducting Al
contact on one end, and a normal Au contact on the other.  The MR at low magnetic field $B$ is quadratic in $B$,
with a characteristic field scale $B_c$ determined by phase coherent paths which encompass not only the wire, but
also the two contacts. 
$B_c$ is essentially temperature independent at low temperatures,  indicating that the area of the phase coherent paths is not
determined by the superconducting coherence length $L_T$ in the normal metal,  which is strongly temperature dependent at low
temperatures.  We identify the relevant length scale as a combination of the electron phase coherence length
$L_\phi$ in the normal metal and the coherence length $\xi_S$ in the superconductor.   
\end{abstract}


\begin{multicols}{2}

The properties of a normal metal (N) in contact with a superconductor (S) have been an active subject of interest in recent
years\cite{NS}.  The proximity of a normal metal with a superconductor induces pair correlations which can extend appreciable
distances into the normal metal. In dirty normal metals (the case of interest here), where  the motion of the electrons is
diffusive, the relevant length scale is the normal metal coherence length $L_T=\sqrt{\hbar D /k_B T}$, where
$D$ is the electronic diffusion coefficient\cite{Belzig}. For typical metallic films, $L_T$ can be as
long as 0.5 $\mu m$ at $T$=1 K.  With modern lithographic techniques, this now represents an experimentally
accessible length scale, and many experiments in the past few years have investigated the proximity effect in
mesoscopic NS structures.  An overview of these experiments can be found in the recent article by Courtois and
Pannetier\cite{Pannetier}.

The microscopic basis of the superconducting proximity effect is Andreev reflection (AR)\cite{Andreev}, whereby an  electron
in the normal metal incident on the NS interface with an energy $\epsilon$ less than the gap $\Delta$ of the superconductor is
reflected as a hole, with the simultaneous generation of a Cooper pair in the superconductor.  AR is a phase coherent process;
the phases of the incident electron and the reflected hole are related through the  macroscopic phase $\phi$ of the
superconductor. This has been elegantly demonstrated by recent interference experiments in so-called Andreev
interferometers\cite{Takayanagi}, which are NS loops where one arm is fabricated from a superconductor, and the other from
a normal metal.  The electrical\cite{AI} and thermoelectric properties\cite{Eom} of these doubly-connected devices have been
found to oscillate periodically  with the magnetic flux coupled to the area of the loop, with a fundamental period
corresponding to one flux quantum $\Phi_0=h/2e$. 

In spite of the tremendous amount of work on the proximity effect, the relevant length $L_p$ that sets the scale for quantum
interference in the proximity regime is not entirely clear. One might argue that $L_p$ should be set by $L_T$, the length
which determines how far superconducting pair correlations can diffuse at a temperature $T$ in the proximity-coupled normal
metal before breaking apart. However, experiments which measure the amplitude of the magnetoresistance (MR) oscillations in
Andreev interferometers indicate that
$L_p$ can be much longer than $L_T$.  For example, Courtois \etal\cite{Courtois} measured the temperature dependence of the MR
oscillations in an Andreev interferometer and found that the oscillation amplitude decreased only gradually with increasing
temperature. According to their analysis, the relevant length scale is $L_T$ with the electron
phase coherence length $L_\phi$ providing an upper cutoff. Only pair correlations with a coherence length greater than the length
$L$ of the normal arm of the Andreev interferometer can contribute to the interference.  The fraction of such correlations is given
by $E_c/k_B T=L_T^2/L^2$, where $E_c=\hbar D/L^2$, giving rise to an oscillation amplitude which decreases with temperature
as $1/T$, in agreement with experiment. 

In this Letter, we describe our measurements on the proximity effect magnetoresistance of short, narrow Au wires.  The wires
are connected on one end to a large Al contact, and on the other to a large Au contact.  The MR is quadratic at low magnetic
fields; however, the characteristic field scale $B_c$ is much smaller than that expected from the dimensions of the wire, but
agrees with what one would expect from contributions of  phase-coherent paths which encompass the wire as well as
the normal \emph{and} superconducting contacts, indicating that the contacts cannot be considered ideal reservoirs. 
Furthermore, at temperatures $T \leq $ 0.5$T_c$, $B_c$ is essentially temperature independent. This indicates that
the relevant length scale for interference is not determined by $L_T$, which varies as
$\sqrt{1/T}$.  At higher temperatures,
$B_c$ \textit{decreases} with increasing temperatures, evidence that the interference is associated with phase
coherent paths whose lengths increase with increasing temperature.  We identify the relevant length for
interference as a combination of $L_\phi$ in the normal metal, and the superconducting coherence length $\xi_S$ in
the superconductor.    

	The samples for this experiment were produced by multi-level electron beam lithography techniques on oxidized silicon
substrates.  Figure 1(a) shows a scanning electron micrograph of the sample discussed in this paper.  It consists of 5 Au
wires of different lengths, each connected to a separate 3$\times$3 $\mu$m$^2$ Au reservoir on one end, and to the same Al
reservoir on the other. All 5 wires are connected to the same  Au strip of width $\simeq$ 0.3 $\mu$m just underneath the Al
film at the Al/Au interface to ensure as much as possible a uniform interface resistance for all samples.  Two additional fine
Au probes on each wire permit us to make four-terminal resistance measurements on the wire by itself, without including any
explicit contributions from the superconductor or the NS interface (see Fig. 1(b)). Additional contacts let us directly
measure the four terminal resistance of the Al film and the NS interface as well. The 50 nm thick Au wires and contacts were
patterned and evaporated first, after which the 80 nm thick Al contact was evaporated following  an Ar$^+$ etch to ensure good
interfaces between the Au and Al films, as evidenced by a measured interface resistance of less than 0.1 $\Omega$.  The area
of the Al contact was 3$\times$20 $\mu$m$^2$, and its transition temperature was
$T_c = 1.24$ K.  In addition to the sample itself, a Au meander wire was fabricated simultaneously to characterize the 
material properties.  From weak localization (WL) measurements on this control sample, $L_{\phi}\cong 3.8$
$\mu$m at $T$ = 35 mK, and the diffusion constant in the Au was determined to be $D \cong 3.0
\times 10^{-4}$ m$^2$/sec, giving $L_T \cong 0.48$ $\mu$m at $T=1$ K.  The samples were measured in a dilution refrigerator using
standard ac lock-in techniques in a magnetic field perpendicular to the sample substrate.  Of the 5 wires shown in Fig. 1(a),
two had one or more contacts disconnected and could not be measured.  The lengths of the remaining three were $L=$1.0, 1.2 and
1.5 $\mu$m, and their widths were $W$ = 120 nm.  

The temperature dependence of the resistance of this and similar samples was studied in detail, and has been
reported elsewhere\cite{Black}. Here we shall concentrate on the low field MR of the proximity coupled Au wires. 
Figure 2 shows the MR of the $L=$1.5 $\mu$m wire at $T=$ 37 mK.  The most noticeable aspect of these data is the
fact that there is not just one MR curve, but a number of distinct curves which are almost identical, except that
they are offset from each other by a magnetic field of $\simeq$1.4 G.  The sample switches spontaneously between
these curves on sweeping the magnetic field.  Similar behavior is observed for the 1.0 and  1.2 $\mu$m wires as
well.  Since this metastability appears only on measurements of the proximity coupled wires below $T_c$ of the Al
film, and not the Au control wire which was measured simultaneously, it is not an experimental artifact, associated
for example with the superconducting solenoid used for the external magnetic field. This hysteretic behavior
appears to be due to metastable screening states in the superconducting contact which correspond to a paramagnetic
response to the external magnetic field. However, this behavior is not the focus of this paper, and will be
discussed in a later publication.  For the remainder of the paper, we will turn our attention to a single MR curve
in order to discuss its magnetic field dependence in detail.  

Figure 3 shows the MR of the $L=$1.5 $\mu$m wire at 35 mK and 1.06 K, along with the MR of the superconducting bank by itself.  At
low magnetic fields, the MR of the wire is quadratic, with the resistance increasing rapidly to its normal state value within
$\simeq$ 7 G.  The superconductor, on the other hand, remains in a resistanceless state until a magnetic field of $\simeq$70
G (at 35 mK), where there is a rapid transition to the normal state resistance.  This indicates that, at low temperatures, the
field scale of the proximity wires is not restricted by the critical field of the superconductor.  We shall now
attempt to understand the magnetic field scale of the MR of the proximity wire.

In analogy with quantum interference effects such as WL in normal metals, the Aharonov-Bohm (AB) phase generated by the
presence of a magnetic field also leads to measurable effects in singly connected structures such as films and wires.  For WL,
which arises from the interference of electrons traversing pairs of time reversed paths, the  application of a magnetic field
destroys this interference and leads to a decrease in resistance\cite{Bergmann}.  An estimate of the characteristic field
required to destroy this interference can be obtained from the physical argument that this  field should correspond to one
flux quantum through the area of the largest possible phase coherent path. For WL, it is $L_\phi$ that determines the length of
the phase coherent paths.  For two dimensional (2D) films, in which both lateral  dimensions (but not the thickness) are larger
than $L_\phi$, the characteristic field is consequently $B_c \sim
\Phi_0 / L_{\phi}^2$.  For one dimensional (1D) wires, where motion in the transverse direction is restricted by
the finite width $W$ of the wire, the corresponding expression is $B_c \sim \Phi_0 /(L_\phi W )$. In the
intermediate case, where a 1D wire of length $L \ll L_\phi$ is connected to 2D probes, the electrons in the wire can
sample regions in the  2D probes in a phase coherence time.  In this case, the characteristic field is not given by
either the 1D or the 2D form, but something in between which depends on the ratio $L/L_\phi$. These
physical arguments are  supported by more quantitative calculations for WL\cite{VenkatPRB}, and have also been
confirmed by experiment\cite{VenkatPRL}.

Drawing on the similarity between the equation of motion for the Cooperon, which determines the WL correction, and the Usadel
equation for the anomalous superconducting Green's function parameter $\Theta$, which determines the proximity effect
corrections in a diffusive normal metal\cite{Belzig}, one would expect a similar situation to occur for the superconducting
proximity effect. Figure 1(c) shows a schematic of a diffusive quasiparticle trajectory in a normal metal wire with a normal
contact on one end, and a superconducting contact on the other. An electron diffusing in the normal metal is Andreev reflected
as a hole at point (1) on the NS interface. The hole picks up an additional phase factor corresponding to the macroscopic
phase $\phi$ of the superconductor.  This hole retraces the trajectory of the incident electron in the opposite direction,
eventually intersecting the NS interface at point (2), where it in turn is Andreev reflected as an electron with the
accumulation of an additional phase factor of -$\phi$.  Since the AR process is phase-coherent, this second  electron, which
retraces the trajectory of the hole, can interfere with the first electron.  In the absence of a magnetic field, the
additional phase shifts introduced by the two AR processes cancel.  To simplify the theoretical analysis, the phase $\phi$ of
the singly-connected superconducting contact is usually assumed  to be a constant along the NS interface, even in the presence
of a magnetic field. In addition, the normal contact is considered to be an ideal `reservoir,' in that $\Theta$ vanishes. 
This is equivalent to a  quasiparticle immediately losing phase memory on entering the normal contact. 

Let us assume for the moment that the relevant phase coherence length $L_p$ in the proximity coupled metal is longer than the
length $L$ of the wire.  If the normal contact is an ideal reservoir, and the superconducting contact has a uniform phase
$\phi$, then the expected field scale is given by $B_c \sim \Phi_0/(LW)$.  Applying this to the $L=$ 1.5 $\mu$m wire whose MR 
is shown in Fig. 3, we obtain an expected field scale of $B_c \sim$ 120 G, more than a factor of 10 larger than the experimentally
observed field scale. This implies that the area of some of the trajectories contributing to the MR are not restricted to the
proximity wire alone, but encompass a much larger area.  For example, if we now relax the assumption that the normal contact is a
phase-randomizing reservoir, one can consider trajectories in which the quasiparticles diffuse coherently into the normal contact
and return to the proximity wire, leading to \emph{nonlocal} contributions to the MR of the wire.  In fact, the
measured $B_c$ corresponds roughly to a phase coherent area comparable to the area of the normal contact. In the absence of a
theory which includes the effects of phase coherence in normal metal contacts, a more quantitative comparison to calculations
based on the quasiclassical theory is difficult.

Further information about $L_p$ can be obtained by investigating $B_c(T)$ as a function of $T$ in the proximity wires. 
Figure 4(a) shows these data for the three proximity wires, as well as the superconducting contact by itself. (Experimentally, we
define $B_c$ as the field at which the extrapolated low field quadratic behavior intersects the saturation value of the MR, as
shown in Fig. 3). $B_c(T)$ reflects the temperature dependence of $L_p$, since $B_c \sim 1/L_p W$
in the 1D wire and $\sim 1/L_p^2$ in the normal contact (2D case), when $L_p$ is shorter than the contact dimensions. At low 
temperatures $B_c$ for all three wires is essentially constant.  At higher temperatures, $B_c$ \emph{decreases} as $T$ increases,
indicating that $L_p$ \textit{increases} as $T$ increases.  Figure 4(b) shows the temperature dependence of $L_\phi$ obtained from WL
measurements on the Au control wire, along with the calculated dependence of $L_T \sim \sqrt{1/T}$. Both lengths
decrease as a function of temperature, exactly opposite to the expected temperature dependence of
$L_p$.  Some understanding of this dependence of $B_c(T)$ for the proximity wires can be obtained by examining  $B_{cS}(T)$ for the
superconducting contact, shown in Fig. 4(a). $B_{cS}(T)$ for a 2D superconductor is determined by the superconducting phase coherence
length
$\xi_S(T)$, i.e., $B_{cS} \sim \Phi_0/ \xi_S^2$ \cite{tinkham}.  The difference for a superconductor, however, is that $\xi_S(T)$
\emph{increases} as $T \rightarrow T_c$, so $B_{cS}$ decreases, as observed experimentally. This suggests that $\xi_S(T)$ also 
plays a role in determining the interference in the proximity coupled normal metal.

Drawing on this information, one can come up with a physical picture for the MR of the proximity wire.  In our discussion
above, we assumed that the phase of the superconducting contact was constant at all points on the NS interface, even in the
presence of a magnetic field.  In reality, however, one should take into account the AB phase that can be accumulated along
the paths in the superconductor.  Figure 1(c) shows an example of one such path.  The AB phase accumulated along the path
will result in an additional contribution to the phase of the interfering quasiparticles corresponding to the magnetic flux
enclosed by the superconducting path, the same physics that gives rise to MR oscillations in Andreev interferometers.  
This means that $B_c$ is now determined by the area of the largest phase coherent trajectory which encompasses both the normal
metal and the superconductor, i.e., a dependence given by something of the form $B_c(\xi_S, L_p) \sim \Phi_0/(\xi_S(T)^2 +
A_{L_p})$, where $A_{L_p}=L_p^2$ if $L_p \gg L$, and $A_{L_p}=L_p W$ if $L_p \le L$.  At low temperatures, $L_p \gg \xi_S$, since
$B_{cS} \gg B_c$.  (This is also in agreement with the fact that $\xi_S(0)$ for Al is about 190 nm.) Hence
$B_c$ is determined  primarily by $L_p$ at low temperatures.  Near $T_c$, however, $\xi_S(T)$ diverges and can be much longer than 
$L_p$, so that $B_c$ is determined essentially by $\xi_S(T)$.  Consequently, $B_c$ for the proximity wires is much smaller than
$B_{cS}$ at low temperatures, but the field scales merge near $T_c$. 

What determines $L_p$, the coherence length in the normal metal?  This question can be answered by
plotting $B_c(\xi_S, L_p)$, with $L_p$ replaced by $L_T$ or $L_\phi$.  In order to compare $B_c(\xi_S, L_p)$ to our
experimental results, we rewrite this function in the form $B_c(\xi_S, L_p) = (1/B_{cS} + A_{L_p}/\Phi_0)^{-1}$.  Since $L_\phi
\gg L$ at all temperatures while $L_T \le L$ above 100 mK, we take $A_{L_\phi}=L_\phi^2$ and $A_{L_T}=L_T W$. 
Figure 4(a) shows the resulting curves, along with the measured $B_c$.  While $B_c(\xi_S,
L_\phi)$ closely follows the experimental curve, $B_c(\xi_S, L_T)$ is  nonmonotonic, showing a maximum of $\simeq$
50 G at a temperature of 300 mK.  This clearly shows that $L_\phi$ is the relevant length scale for
interference in the normal metal, not $L_T$.          

In conclusion, our results show the relevant length scale that determines Aharonov-Bohm type interference in proximity coupled
normal metals is the electron phase coherence length $L_\phi$.  A detailed quantitative analysis of the MR needs to take into
account the contributions of nonlocal phase coherent transport in both the superconducting and normal contacts of a proximity
effect device.  

This work was supported by the NSF through DMR-9801982, and by the David and Lucile Packard Foundation.

\end{multicols}

\vspace{4cm}
\begin{figure}[p]
\caption{(a) Scanning electron micrograph of sample: the light areas are normal metal (Au) and the dark
rectangular is superconductor (Al). (b) Schematic of one of the wires with current (I) and voltage (V) 
probes. (c) Schematic representation of Andreev reflection and a quasiparticle 
trajectory in vicinity of the NS interface.  See text for details}
\end{figure}

\begin{figure}[p]
\caption{Resistance versus magnetic field for the 1.5 $\mu$m long proximity wire at 37 mK. Vertical solid lines
indicate the points of MR jumps from one branch to another. The three dashed curves are quadratic fits to the low
field behavior.}
\end{figure}

\begin{figure}[p]
\caption{Resistance versus magnetic field for the 1.5 $\mu$m wire and
superconducting film at $T=$ 35 mK (curves 1) and $T=$ 1.06 K (curves 2). Left axis: 1.5 $\mu$m wire, open symbols;
right axis: superconducting film, solid line.  Probe configuration for proximity wire: current I$_1$-I$_{3,4}$,
voltage V$_1$-V$_2$;  for superconductor: current I$_3$-I$_4$, voltage V$_3$-V$_4$.  The values of the
characteristic fields $B_c$ and $B_{cS}$ are shown.}
\end{figure}

\begin{figure}[p]
\caption{(a) Temperature dependences of $B_{cS}$ ($\circ$) and 
$B_c$ for the 1.5 ($\Diamond$), 1.2 ($\triangle$) and 1.0 ($\Box$) $\mu$m long wires. Calculated $B_c(T)$ for
$L_p=L_T$ (dashed curve) and $L_p=L_\phi$ (dotted curve), as described in the text. The solid line is a guide to the eye. 
(b) Measured temperature dependence of $L_\phi$, and calculated dependence of $L_T$. 
The temperature dependence of $\xi_S$ near $T_c$ is also shown.}
\end{figure}

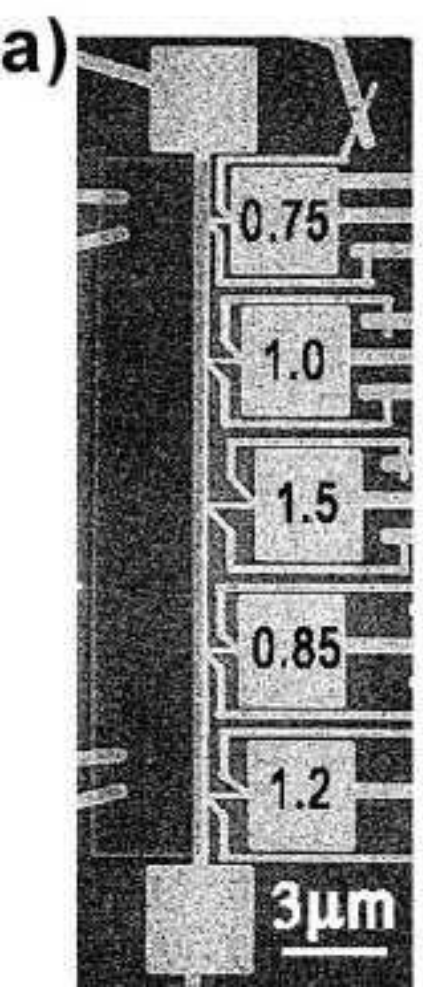
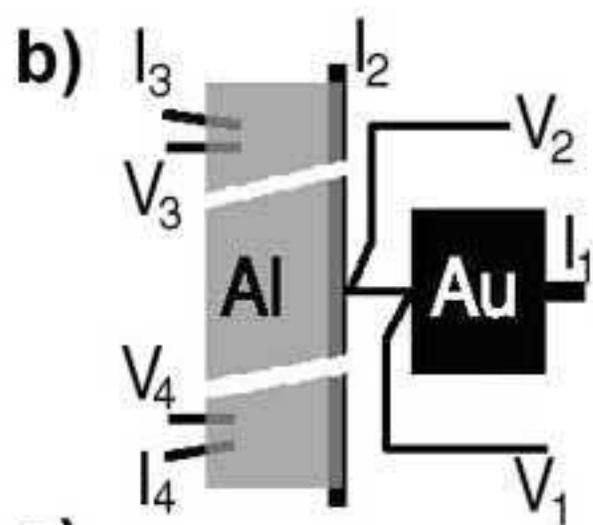
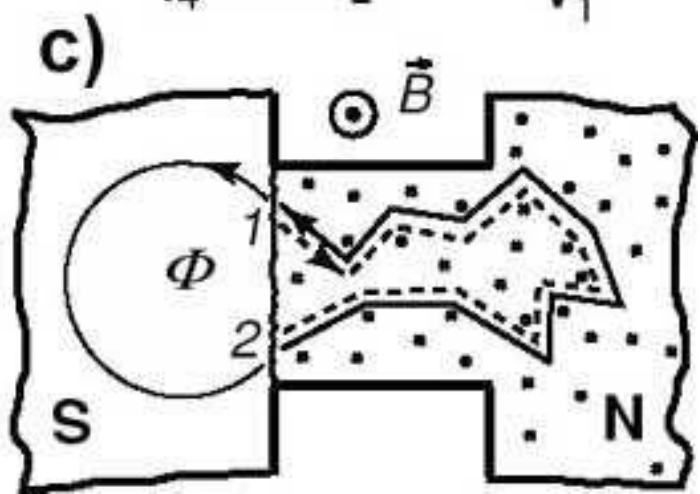
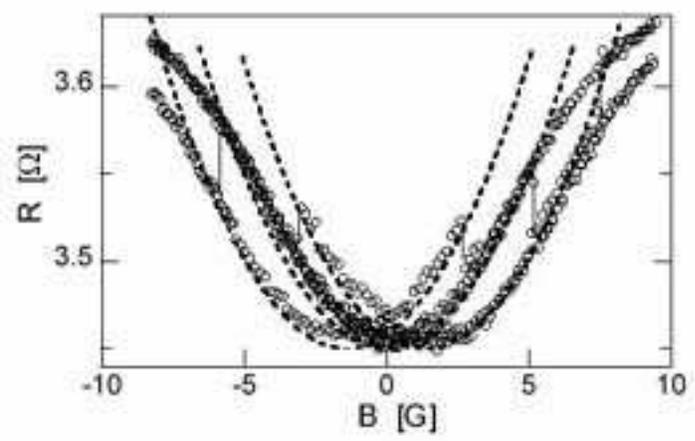
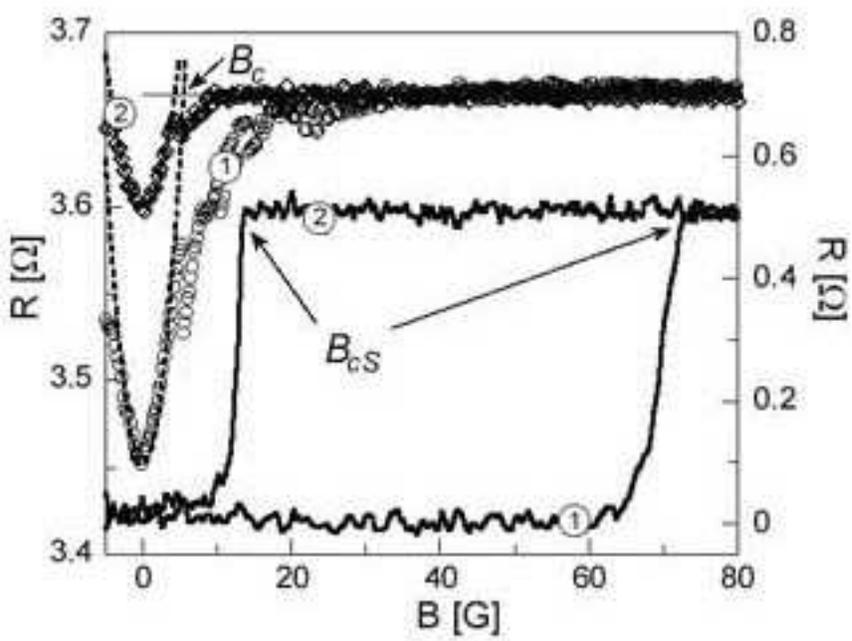
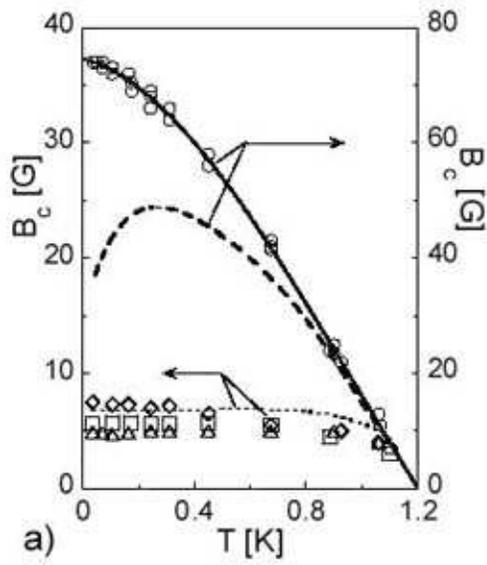
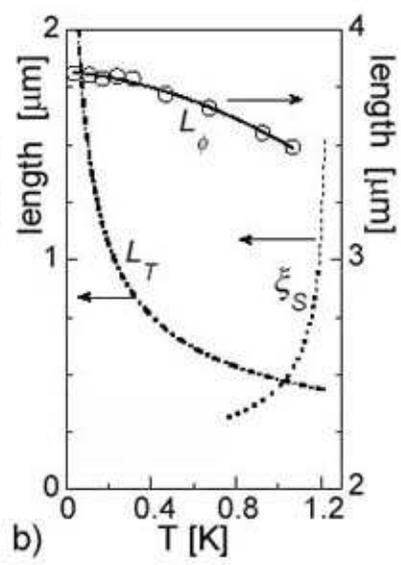
\end{document}